\documentstyle[psfig]{mn}
\newif\ifAMStwofonts

\def\p#1#2{{\partial#1\over\partial#2}}

\ifoldfss
  \ifCUPmtlplainloaded \else
    \NewTextAlphabet{textbfit} {cmbxti10} {}
    \NewTextAlphabet{textbfss} {cmssbx10} {}
    \NewMathAlphabet{mathbfit} {cmbxti10} {} 
    \NewMathAlphabet{mathbfss} {cmssbx10} {} 
  \fi
  \ifAMStwofonts
    \ifCUPmtlplainloaded \else
      \NewSymbolFont{upmath} {eurm10}
      \NewSymbolFont{AMSa} {msam10}
      \NewMathSymbol{\upi}     {0}{upmath}{19}
      \NewMathSymbol{\umu}     {0}{upmath}{16}
      \NewMathSymbol{\upartial}{0}{upmath}{40}
      \NewMathSymbol{\leqslant}{3}{AMSa}{36}
      \NewMathSymbol{\geqslant}{3}{AMSa}{3E}

       \let\le=\leqslant
       
    \fi
  \fi
\fi 

\ifnfssone
  \newmathalphabet{\mathit}
  \addtoversion{normal}{\mathit}{cmr}{m}{it}
  \addtoversion{bold}{\mathit}{cmr}{bx}{it}
  \newmathalphabet{\mathbfit} 
  \addtoversion{normal}{\mathbfit}{cmr}{bx}{it}
  \addtoversion{bold}{\mathbfit}{cmr}{bx}{it}
  \newmathalphabet{\mathbfss} 
  \addtoversion{normal}{\mathbfss}{cmss}{bx}{n}
  \addtoversion{bold}{\mathbfss}{cmss}{bx}{n}
  \ifAMStwofonts
    \ifCUPmtlplainloaded \else
      \UseAMStwoboldmath
      \makeatletter
      \new@mathgroup\upmath@group
      \define@mathgroup\mv@normal\upmath@group{eur}{m}{n}
      \define@mathgroup\mv@bold\upmath@group{eur}{b}{n}
      \edef\UPM{\hexnumber\upmath@group}
      \new@mathgroup\amsa@group
      \define@mathgroup\mv@normal\amsa@group{msa}{m}{n}
      \define@mathgroup\mv@bold\amsa@group{msa}{m}{n}
      \edef\AMSa{\hexnumber\amsa@group}
      \makeatother
      \mathchardef\upi="0\UPM19
      \mathchardef\umu="0\UPM16
      \mathchardef\upartial="0\UPM40
      \mathchardef\leqslant="3\AMSa36
      \mathchardef\geqslant="3\AMSa3E

       \let\le=\leqslant

    \fi
  \fi
\fi 

\ifnfsstwo
  \DeclareMathAlphabet{\mathbfit}{OT1}{cmr}{bx}{it}
  \SetMathAlphabet\mathbfit{bold}{OT1}{cmr}{bx}{it}
  \DeclareMathAlphabet{\mathbfss}{OT1}{cmss}{bx}{n}
  \SetMathAlphabet\mathbfss{bold}{OT1}{cmss}{bx}{n}
  \ifAMStwofonts
    \ifCUPmtlplainloaded \else
      \DeclareSymbolFont{UPM}{U}{eur}{m}{n}
      \SetSymbolFont{UPM}{bold}{U}{eur}{b}{n}
      \DeclareSymbolFont{AMSa}{U}{msa}{m}{n}
      \DeclareMathSymbol{\upi}{0}{UPM}{"19}
      \DeclareMathSymbol{\umu}{0}{UPM}{"16}
      \DeclareMathSymbol{\upartial}{0}{UPM}{"40}
      \DeclareMathSymbol{\leqslant}{3}{AMSa}{"36}
      \DeclareMathSymbol{\geqslant}{3}{AMSa}{"3E}

       \let\le=\leqslant

    \fi
  \fi
\fi 

\ifCUPmtlplainloaded \else
  \ifAMStwofonts \else 
    \def\upi{\pi}
    \def\umu{\mu}
    \def\upartial{\partial}
  \fi
\fi

\title{On the Fate of Gas Accreting at a Low Rate onto a Black Hole}
\author[Roger D. Blandford and Mitchell C. Begelman]
       {Roger D. Blandford$^1$\thanks{E-mail: rdb@tapir.caltech.edu} and Mitchell C. Begelman$^2$\thanks{E-mail: mitch@jila.colorado.edu. Also at Department of Astrophysical and Planetary Sciences, University of Colorado}\\
        $^1$Theoretical Astrophysics, Caltech 130-33, Pasadena, CA 91125, USA\\
        $^2$JILA, University of Colorado, Boulder, CO 80309-0440, USA}
\date{Accepted 1998.
      Received 1998}
\pagerange{\pageref{firstpage}--\pageref{lastpage}}
\pubyear{1998}
\begin{document}
\maketitle
\label{firstpage}
\begin{abstract}
Gas supplied conservatively 
to a black hole at rates well below the Eddington rate may not be
able to radiate effectively and the net energy flux, including the
energy transported by the viscous torque, is likely to be close to zero
at all radii. This has the consequence that the gas accretes with 
positive energy so that it may escape. Accordingly, we propose that
only a small fraction of the gas supplied actually falls onto the black hole 
and that the binding energy it releases is 
transported radially outward by the torque so as to drive away 
the remainder in the form of a wind. This is a generalization of
and an alternative to an ``ADAF" solution. Some observational 
implications and possible ways to distinguish these two types of 
flow are briefly discussed. 
\end{abstract}
\section{Introduction}
It has often been supposed that gas that accretes onto a black hole 
radiates liberated binding energy with an efficiency of 
$\sim0.1c^2\sim10^{20}$~erg g$^{-1}$.
This is not always so, as exemplified by 
observations of the black hole in our Galactic center,
where it appears that gas is supplied at a rate that has 
been estimated to lie in the range $\sim10^{20-22}$~g s$^{-1}$  
\cite{fm97}, while the bolometric luminosity 
is found to be $\sim10^{36-37}$~erg s$^{-1}$ 
\cite{mah98}. Consequently,
the radiative efficiency could be as low as 
$\sim10^{14}$~erg g$^{-1}\sim10^{-7}c^2$ and is unlikely to be 
more than $\sim10^{-4}c^2$. 

If gas falls onto a black hole from a few gravitational
radii via a thick disk, with shear stress per unit pressure 
$\alpha$, then the fraction of the energy of 
an individual hot ion that is transferred by Coulomb scattering 
to the electrons (which are almost solely responsible 
for the radiation) is $f_e\sim(\dot M/\dot M_E)\alpha^{-2}$,
where $\dot M_E=L_E/c^2$ is the Eddington accretion rate. 
Therefore, if (i) viscous dissipation heats only the ions, (ii)
the ions couple to the electrons only through Coulomb scattering, and
(iii) $\alpha\ga0.1$, then the radiative efficiency is 
plausibly low enough to account for the observations of the Galactic center.
The plausibility of condition (i) has been argued recently by Gruzinov
\shortcite{gruz98} and Quataert \shortcite{quat98} (but see also
Blackman 1998, Bisnovatyi-Kogan \& Lovelace 1997),  
(ii) seems reasonable in the absence 
of a specific proposal for non-Coulombic heating, and (iii) can be a feature 
of an Advection-Dominated Accretion Flow, or ADAF, in which gas 
accretes quasi-spherically onto a black hole carrying a
large amount of internal energy across the horizon
(Narayan \& Yi 1994; Kato, Fukue, \& Mineshige 
1998, and references therein).

However, as we discuss below, the gas in ADAF solutions 
appears to be generically unbound.  We therefore propose in this letter that
ADAF solutions be modified to include a powerful wind that
carries away mass, angular momentum, and energy from the accreting gas. 
We describe a family of solutions where the rate at which 
gas is swallowed by the black hole is only a tiny fraction of the 
rate at which it is supplied, and where, in the limiting
case, the binding energy of a gram of gas at a few gravitational 
radii drives off a kilogram of gas 
from a few thousand gravitational radii. Disk-wind solutions 
based on quite different principles have also been 
proposed recently by Xu \& Chen (1997) and Das (1998).
\section{Fundamentals of Accretion Theory}
First, we review some principles.
Consider thin disk accretion with angular velocity $\Omega$, 
inflow speed $v \ll \Omega r$, mass per unit radius $\mu$ and 
specific angular momentum $\ell$.
In assuming that the disk is thin, we are implicitly 
supposing that the gas can remain cold by radiating away its internal energy.
Let the torque that the disk interior to radius $r$ exerts 
upon the exterior disk be $G(r)$. The equations of mass and 
angular momentum conservation are then
\begin{equation}
\label{pmut}
\p\mu t=\p{\mu v}r;\qquad\p{\mu\ell}t=\p{\mu v\ell}r-\p Gr ,
\end{equation}
e.g.,  Kato et al. (1998).  
These equations immediately imply
\begin{equation}
\label{pgr}
\p Gr={\mu v\ell\over2r};\qquad\p\mu t=2\p{}rr^{1/2}\p Gr
\end{equation}
where we have assumed the Keplerian relation 
$\ell=r^{1/2}$ and set $GM=c=1$ \cite{lbp74}. 

We can combine equations (\ref{pmut}) to obtain an energy equation
\begin{equation}
\label{pmuet}
\p{\mu e}t+\p{(\Omega G -\mu v e)}r=G\p\Omega r
\end{equation}
where $e=-\Omega\ell/2$ is the Keplerian binding energy, the sum of
the kinetic and potential energy per unit mass.
(Note the presence of a contribution to the energy flux from the rate at which the
torque, $G$, does work on the exterior disk.) The right-hand side represents a radiative loss
of energy. Evaluating it, we recover the standard result
that the local radiative flux, in a stationary disk,
is three times the rate of local loss of binding 
energy (D.~Lynden-Bell, K.~Thorne, quoted in Pringle \& Rees 1972).

Next consider the opposite limiting case when the gas cannot cool and there is no extraneous
source or sink of energy. Adding thermodynamic terms to the energy equation, we obtain
\begin{equation}
\label{pmueut}
\p{\mu(e+u)}t+\p{(\Omega G -\mu v(e+h))}r=G\p\Omega r+\mu T{ds\over dt}
\end{equation}
where $u$ is the vertically-averaged internal energy density, 
$h$ is the enthalpy density, and $s$ is the entropy
density \cite{ll59}.
As there are no sources or sinks of energy, the right-hand side must vanish:
\begin{equation}
\label{muts}
\mu T{ds\over dt}=T\left[\p{\mu s}t-\p{\mu vs}r\right]=-G\p\Omega r.
\end{equation}
As the gas has pressure, we must also satisfy the radial equation 
of motion:
\begin{equation}
\label{pvt}
\p vt-v\p vr+\Omega^2r={1\over r^2}+{1\over\rho}\p Pr.
\end{equation}
\section{Advection-Dominated Accretion Flows}
The basic idea and assumptions are set out most transparently in 
Narayan \& Yi (1994; cf. also Ichimaru 1977, Abramowicz et al. 1995, Narayan \& Yi 1995). In the 
simplest, limiting case, it is assumed 
that there is a stationary, one-dimensional, self-similar flow 
of gas with $\mu\propto r^{1/2}$,
$\Omega\propto r^{-3/2}$, and $v,a\propto r^{-1/2}$, where 
$a=[(\gamma-1)h/\gamma]^{1/2}$ is the isothermal sound speed
and the radial velocity $v \ll \Omega r$.
The requirement that $P\propto r^{-5/2}$ transforms the radial equation 
of motion into
\begin{equation}
\label{omeg}
\Omega^2r^2-{1\over r}+{5a^2\over2}=0.
\end{equation}
Conservation of mass, angular momentum and energy gives
\begin{equation}
\label{dotm}
\mu v\equiv\dot m={\rm constant}
\end{equation}
\begin{equation}
\label{dotg}
\dot mr^2\Omega-G=F_\ell
\end{equation}
\begin{equation}
\label{dote}
G\Omega-\dot m\left[{1\over2}\Omega^2r^2-{1\over r}+{\gamma a^2\over
\gamma-1}\right]=F_E
\end{equation}
where the inwardly directed angular momentum flux, $F_\ell$, and the outwardly directed
energy flux, $F_E$, are constant if there are no sources and sinks of angular momentum
or energy. Now, the terms on the left-hand side of equation (\ref{dotg})
scale $\propto r^{1/2}$ and those of equation (\ref{dote}) scale $\propto r^{-1}$.
Therefore, if we require the flow to be self-similar over several decades of radius, 
both constants must nearly vanish.  In the limit,
$F_\ell =F_E=0$.  

Combining equations, we derive expressions for the sound speed $a$ and
the Bernoulli constant $Be$:
\begin{equation}
\label{heq}
a^2=\left[{3(\gamma-1)\over5-3\gamma}\right]\Omega^2r^2={6 (\gamma-1)\over(9\gamma-5)r}
\end{equation}
\begin{equation}
\label{bern}
Be\equiv{1\over2}\Omega^2r^2-{1\over r}+{\gamma a^2\over \gamma-1 }=\Omega^2r^2 .
\end{equation}
The elementary ADAF solution is then completed by defining 
an $\alpha$ viscosity parameter through, e.g.,  $G=\dot mr^2\Omega=\alpha\mu ra^2$,
which then implies $v={\alpha a^2 /\Omega r }$, 
assuming that $\alpha \ll (5/3-\gamma)^{1/2}$. (Note that this, conventional, definition of $\alpha$ differs slightly from the Newtonian prescription used by Narayan \& Yi 1994.)

This solution has some features (as noted by Narayan \& Yi 1994) that make it somewhat 
problematic.  The first is a technical, though somewhat subtle point.
As can be seen from equation (\ref{heq}), $\gamma=5/3$ is a singular case and, if imposed strictly, requires the flow to be non-rotating.
This is familiar from the Bondi (1952) analysis which found a self-similar
non-rotating inflow only when $\gamma=5/3$.  Narayan \& Yi (1994) avoid 
this problem by supposing that the magnetic energy density
is comparable with the ion energy density and behaves dynamically like a $\gamma=4/3$ 
gas when it is highly turbulent so that a composite specific heat ratio of $\gamma=3/2$
is appropriate. But if the magnetic energy density is maintained well below equipartition values,
as numerical simulations of shearing flows suggest is the case (e.g., Balbus \& Hawley 1998), then the internal energy must be dominated by the non-relativistic
ions, $\gamma$ is very close to $5/3$, and $\Omega^2 \approx (5-3\gamma)/5r^3$, well below the Keplerian value. To match onto the ADAF solutions, weakly magnetized flows would have to lose
most of their angular momentum at large radii in a manner likely to 
unbind much of the gas.  Note also that for slowly rotating weakly magnetized
flows, the $\alpha$ prescription is inappropriate and the small differential
rotation makes the generation of field less likely.

The second concern is more fundamental.
The Bernoulli constant, $Be$ (equation [\ref{bern}]), is necessarily positive.  
This implies that any exposed gas can escape to infinity 
with positive energy. Furthermore, the value of $Be$ increases as $\gamma$
decreases from $5/3$, hence addressing the first problem by including
an equipartition magnetic energy density would exacerbate this difficulty.
These problems are not simply a consequence of assuming self-similarity but
stem from the fact that torque transports energy as well as 
angular momentum and that, in a steady state, the angular momentum and
energy fluxes are conserved.  Provided that the mechanical and torque
contributions to the angular momentum (energy) flux are separately 
increasing (decreasing) functions of $r$, while their sums assume the constant values
$F_\ell=O(r_{\rm in}^{1/2})$ ($F_E=O(r_{\rm tr}^{-1})$) in terms of the inner (outer) 
radius $r_{\rm in}$ ($r_{\rm tr}$), we deduce that $F_\ell$ ($F_E$) must be relatively close to zero at intermediate radii, $r$, where $r_{\rm in}\ll r \ll r_{\rm  tr}$. Equation (\ref{bern}) then follows
without using equation (\ref{omeg}) which is where self-similarity
is introduced. Therefore $Be\sim\Omega^2r^2$ at intermediate radii
because as much energy has to be transported outward by the torque
as inward by the mass. (By contrast, there is a net radial inflow of entropy.) 

Thirdly, in an elaboration upon this model, it is supposed that 
a conical velocity field extends to the polar axis, at least for large
$\alpha$ (Narayan \& Yi 1995; Narayan, Kato, \& Honma 1997).  However, in this solution, the gas is in hydrostatic
equilibrium along the axis but is unsupported at its base and it seems hard
to avoid the formation of a funnel from which gas can escape. 

These considerations motivate us to investigate flows in which powerful winds
carry off enough of the mass, angular momentum and energy to bind the gas to the
hole and to allow accretion to proceed.
\section{Advection-Dominated Inflow-Outflow Solutions}
We continue to assume that the accreting gas cannot cool, that 
$v \ll \Omega r$ and that the ions dominate the equation of state so that
$\gamma=5/3$. (As with ADAFs, generalization
to relax each of these assumptions is straightforward.)  
Let the mass inflow rate satisfy
\begin{equation}
\label{wdotm}
\dot m\propto r^p;\qquad0\le p<1 .
\end{equation}
(The restriction on the exponent $p$ allows the accreting mass to decrease 
with decreasing radius, while the energy released can still increase.)

The inward flow of angular momentum satisfies
\begin{equation}
\label{wfl}
F_\ell=(\dot mr^2\Omega-G)=\lambda\dot mr^{1/2};\qquad\lambda>0.
\end{equation}
Similarly, the outward flow of energy is 
\begin{equation}
\label{wfe}
F_E=G\Omega-\dot m\left({1\over2}\Omega^2r^2-
{1\over r}+{5a^2\over2}\right)={\epsilon\dot m\over r};\qquad\epsilon>0.
\end{equation}
Equivalently, for the specific
angular momentum and energy carried off by the wind, we have
\begin{equation}
\label{wdfl}
{dF_\ell\over d\dot m}={\lambda(p+1/2)r^{1/2}\over p};\qquad
{dF_E\over d\dot m}={\epsilon(p-1)\over pr}.
\end{equation}

\begin{figure*}
\psfig{figure=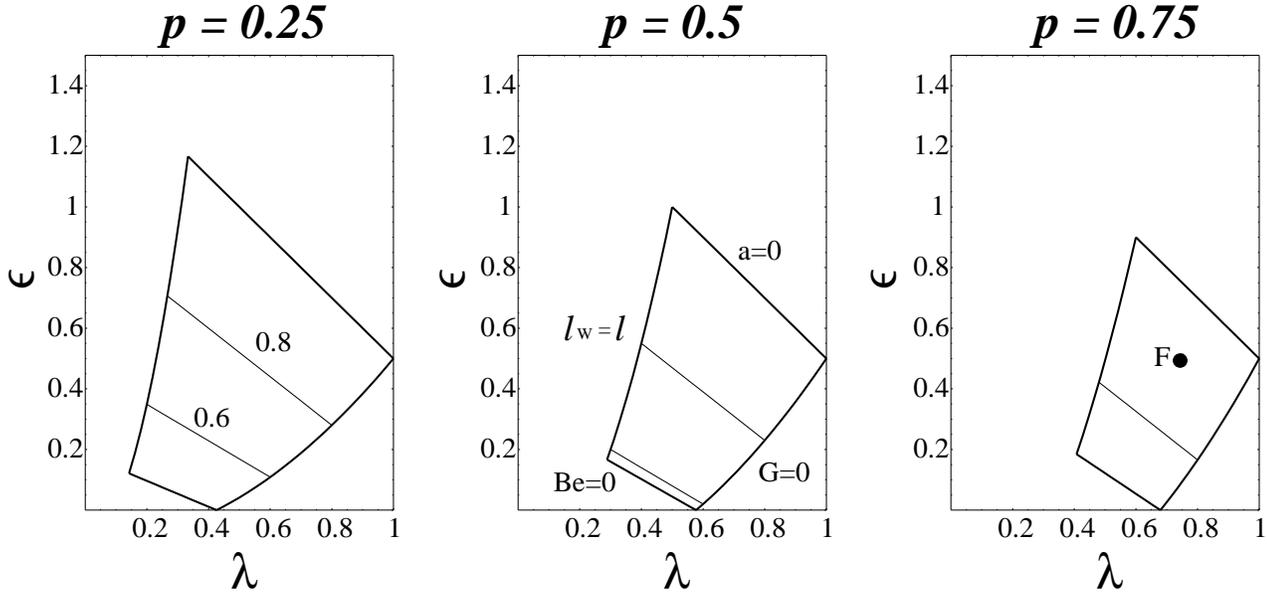,width=0.95\textwidth}
\caption{Allowed regions and limiting cases in the $\lambda-\epsilon$ plane for three
values of the mass loss exponent $p$.  The allowed region, shown with bold
lines, is defined by the following
four constraints: (i) $a,H>0$, (ii) $G>0$, (iii) $\ell_W>\ell$, (iv) $Be<0$. (See text.) 
The light lines correspond to angular velocities equal to 0.6, 0.8 times the Keplerian angular velocity.  The point $F$ for $p=0.75$ corresponds to the fiducial solution.}
\end{figure*}

We use the ADAF self-similar scalings except that we must allow for mass
loss.  The radial equation of motion now gives
\begin{equation}
\label{wom}
\Omega^2r^2-{1\over r}+(5/2-p)a^2=0.
\end{equation}
Similarly, the Bernoulli constant becomes
\begin{equation}
\label{wbe}
Be={\Omega^2r^2\over2}-{1\over r}+{5a^2\over2}
=pa^2-{1\over2}\Omega^2r^2
\end{equation}
and it can have either sign. Combining these equations, we obtain
\[
\Omega r^{3/2}={(5-2p)\lambda\over 15-2p }
\]
\begin{equation}
\label{womeg}
+\quad{[(5-2p)^2\lambda^2+(15-2p)(10\epsilon+4p-4\epsilon p)]^{1/2}
\over 15-2p};
\end{equation}
\[
\left({H\over r}\right)^2=ra^2={2\lambda(5-2p)+6-4\epsilon\over15-2p}
\]
\begin{equation}
\label{wh}
+\quad{2\lambda
[(5-2p)^2\lambda^2+(15-2p)(10\epsilon+4p-4p\epsilon)]^{1/2}\over15-2p} .
\end{equation}
As $a^2$ must be positive, we conclude that $\epsilon<3/2-\lambda$.
The torque is now given by 
$G=\dot m(r^2\Omega-\lambda r^{1/2})=\alpha\mu ra^2$.
As $G>0$, we find that  
\begin{equation}
\epsilon>{(5+2p)\lambda^2-4p\over10-4p}.
\end{equation}
Finally, solving for the inflow speed, we obtain
$v=\alpha r^{1/2}a^2/(\Omega r^{3/2}-\lambda)$.

Even given our simplifying assumptions, there are 
three independent, adjustable parameters,
$p,\lambda,\epsilon$, that depend upon the details of the wind (Fig.~1). 
Let us consider some special cases.
\begin{enumerate}
\item $p=\lambda=\epsilon=0$. There is no wind and the system reduces to the non-rotating
Bondi solution.
\item $p=\lambda=0$, $\epsilon=3(1-f)/2$. This corresponds to flow with no wind but with radiative loss, which carries away energy but not angular momentum. The parameter $f$, introduced by Narayan \& Yi (1994), is defined by the relation $\dot mTdS/dr=fGd\Omega/d
r$.
\item $p=0$, $\lambda=1$, $\epsilon=1/2$. This describes a magnetically-dominated 
wind with mass flow conserved in the disk. All of the angular 
momentum and energy is carried off by a wind with $dF_E/dF_\ell=
\Omega$ (cf. Blandford \& Payne 1982, K\"onigl 1991). 
There is no dissipation in the disk, which is cold and thin. 
\item $\lambda=2p[(10\epsilon+4p-4\epsilon p)
/(2p+1)(4p^2+8p+15)]^{1/2}$. This corresponds to 
a gasdynamical wind where $dF_\ell/d\dot m\equiv\ell_W=r^2\Omega\equiv\ell$. 
The wind carries off its own angular
momentum at the point of launching and does not exert any reaction torque on the remaining gas in the disk. Any magnetic coupling to the
disk implies $\ell_W>\ell$.
\item $ra^2=r^3\Omega^2/2p=1/(p+5/2)$. This corresponds to a 
marginally bound flow with vanishing Bernoulli constant (Fig.~2). In practice, it is expected that $Be<0$.
\item $p=0.75$, $\lambda=0.75$, $\epsilon=0.5$. This is an intermediate solution,
with $Be=-0.35/r$, that carries off
the specific angular momentum of the disk and has a 
velocity at infinity of $0.41$ times the escape velocity 
from the point of origin. The angular velocity is 90 percent of the Keplerian value,
the disk thickness is $H \sim0.3r$ and the 
inflow speed is $v=0.56\alpha$ times the Kepler velocity. 
Only a fraction $(r_{{\rm in}}/r_{{\rm tr}})^{3/4}$ 
of the mass supplied will reach the hole.
\end{enumerate}
\section{Discussion}
The application of our advection-dominated inflow-outflow solutions
(``ADIOS'') to describe real astrophysical flows depends upon
several considerations.  Firstly, we have assumed that the
viscosity is primarily hydromagnetic and dissipates most of the energy locally 
into the ions and that electron heating is ineffective.  
If there is efficient electron heating, then neither ADAFs nor these
ADIOS flows are likely to be of much relevance.  Alternatively, it
is possible that the rate of local dissipation is not given by
$-Gd\Omega/dr$; instead the energy may be transported away by large-scale 
magnetic field, which can also drive an outflow. Secondly, we 
have taken the numerical simulations of MHD shear flows at face value
and supposed that $\alpha\sim0.01$. 
If $\alpha>0.1$, then the radial kinetic energy 
must be included. Just as with the ADAF solutions, this does not change 
their character. Presumably, simulations designed to mimic more closely 
ADAFs or ADIOS flows are possible and might determine the level to which the 
field energy density can grow.
Thirdly, we assume that there is some means for launching an orderly
wind from exposed surfaces that drains energy away from the interior
of the accretion flow. We expect that the wind will be hydromagnetic and will 
extract angular momentum as well as energy, just as in the
solar wind.  However, pure thermal winds are also possible,
especially as ADAFs are probably H\o iland unstable (Begelman \& Meier 1982, Narayan \& Yi 1994). The resulting convection will further increase $Be$ at high latitude.

ADIOS models can be elaborated in much the same way as the ADAF
solutions. The influence of boundary conditions on the similarity solutions
can be followed by directly 
integrating the equations of motion \cite{kato98}. General relativity
has been included at the inner boundary for ADAFs (Abramowicz et al. 1996, 
Igumenshchev \& Beloborodov 1997, Popham \& Gammie 1998) and this approach can be applied to ADIOS as well. If one adopts the lower values of $\alpha$ advocated here, the flow around the black hole may look rather similar to the ion torus model of Rees et al. (1982), although the flow may turn out to be non-stationary. 

\begin{figure}
\psfig{figure=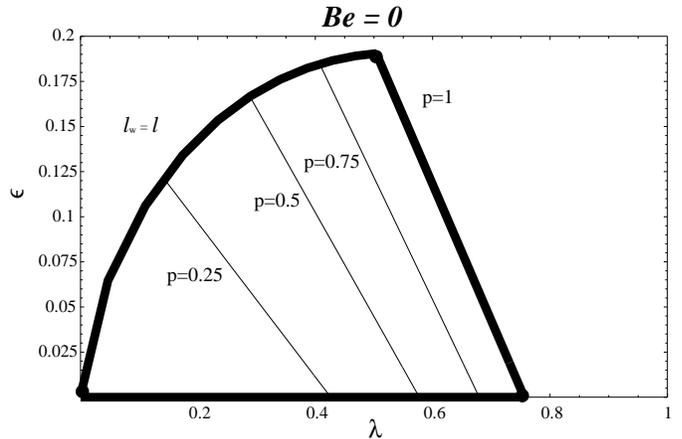,width=0.5\textwidth}
\caption{Accretion flows with vanishing Bernoulli constant. 
The allowed region in the $\lambda-\epsilon$ plane is as defined in 
Fig.~1.  The contours are of constant
mass loss exponent $p$.  Flows with $Be<0$ carry away more energy.}
\end{figure}

Perhaps the most careful application of an ADAF is to our
own Galactic center \cite{mah98}.  The central black hole 
mass is measured to be
$2.6\times10^6$~M$_\odot$ \cite{eck97}
and the detailed ADAF model has a steady mass inflow of
$5\times10^{20}$~g s$^{-1}$ and extends over five decades of radius.  
The model spectrum passes through the mm and X-ray observational
data, complies with upper limits in the infrared, and slightly underestimates the
radio emission.  To illustrate the changes caused by 
substituting our fiducial ADIOS for the ADAF model, suppose
that the flow extends from $\sim3\times10^4m = 10^{16}$~cm
to $\sim3m$ and that the mass supply
rate is as high as $\sim5\times10^{21}$~g s$^{-1}$.  The mass accretion rate onto
the hole is then only $\sim5\times10^{18}$~g s$^{-1}$, significantly lower than in the 
ADAF solution despite the fact that we have assumed a mass supply 
10 times higher.  If $\alpha\sim0.01$, the ions will not cool
and the density close to the hole will be similar to that 
adopted in the ADAF solution, so that a somewhat similar
spectrum can be created, while the mass of the hole increases at a much smaller rate. 

The winds themselves may be sources of observable emission,
especially when they pass through a terminal shock.  Furthermore, the outflows, 
with or without magnetic field, can be self-collimating 
and form jets. (We note that the outflow in M87 appears to be partially
collimated within 60 gravitational radii, cf.~Junor \& Biretta 1995).

The ADAF model has also been applied to black hole X-ray binaries 
and their various spectral states, most notably the ``quiescent'',
``low'' and ``high'' states, have been interpreted as a sequence of flows with 
increasing $\dot M$ (Esin, McClintock, \& Narayan 1997). 
These models can account for the luminosities at which these transitions
occur only if 
the viscosity is high, $\alpha\sim0.3$, and the same would be true for an ADIOS.

One application of the ADIOS model, that may lead to 
a clean observational test, is to neutron star accretion.  Radiatively 
inefficient flow onto the surface of a neutron star is not possible 
for a conventional ADAF solution, because there is inevitably a large release
of energy with efficiency $\sim10^{20}$~erg g$^{-1}$ at the surface.  
However, with an ADIOS, the central density of the gas can be 
greatly reduced relative to an ADAF with the same mass supply at large radius
and it is still possible to have a flow with low radiative efficiency. 
Numerical computations of the emergent spectrum will be necessary 
to see if these flows can model neutron star accretion in low states.

Finally, we note that there are many similarities between accretion at high
rates and low rates. In the former case, the radiative efficiency
is low because electron scattering traps the radiation \cite{beg82}.
These flows also have to lose excess energy and angular momentum,
and winds, like those observed in Broad Absorption Line Quasars, 
provide one way by which this may be accomplished.  
Radiation-dominated analogues of ADIOS exist (Blandford \& Begelman, 
in preparation), and may be relevant for this case.
\section*{Acknowledgments} 
This work was supported in part by NSF grants AST 95-29170 and AST 95-29175,
and NASA grants NAG 5-2837 and NAG5-7007. We thank Martin Rees for 
his encouragement and insights, acknowledge useful discussions with 
Phil Maloney, Mike Nowak, Chris Reynolds and J\"orn Wilms and constructive 
comments from Marek Abramowicz, Jean-Pierre Lasota, Ramesh Narayan
and the referee. RB thanks the Max-Planck-Institut f\"ur Astrophysik 
for hospitality during the completion of this research.

\bsp
\label{lastpage}
\end{document}